\documentclass[%
 reprint,
 amsmath,amssymb,
 aps,
]{revtex4-2}

\usepackage{float}
\usepackage{graphicx}
\usepackage{dcolumn}
\usepackage{bm}

\usepackage{hyperref}

\begin{document}

\title{Intrinsic Curvature Controls Spontaneous Entanglement
of Elastic Rods}

\author{Fani Derveni}%
\email{fani.derveni@cornell.edu}
\affiliation{%
 Scalable MetaStructures Laboratory, Cornell University; \\
 Flexible Structures Laboratory, Ecole Polytechnique Fédérale de Lausanne (EPFL)
}%

\author{Javier Sabater}
\affiliation{%
Flexible Structures Laboratory, Ecole Polytechnique Fédérale de Lausanne (EPFL)
}

\author{Pedro M. Reis}%
\affiliation{%
   Flexible Structures Laboratory, Ecole Polytechnique Fédérale de Lausanne (EPFL)
}%


\begin{abstract}
\noindent We report results from an experimental investigation on the role
of intrinsic curvature in the spontaneous entanglement of elastic
rods subjected to controlled tumbling. We examine two distinct
entanglement mechanisms, knotting and locking, in a statistical
ensemble of straight and helical rods with varying arc length and
natural curvature. Our primary finding is that natural curvature
exerts a dominant, non-monotonic influence on knotting, with a
critical curvature that suppresses knot formation and significantly
reduces the probability of knotting relative to equivalent straight
rods. In contrast, locking is largely insensitive to curvature and
scales primarily with rod length. These results, obtained for the
macroscopic experimental system studied here, identify geometric
parameters that govern distinct entanglement modes in tumbled
elastic rods and may inform entanglement processes in microscopic
filamentary systems such as DNA.
\end{abstract}

\maketitle

The spontaneous entanglement of filaments is a problem that most people have encountered firsthand: headphone cables pulled from a pocket, garden hoses coiled in a shed, or telephone cords left on a desk inevitably form tangles that resist straightening. Beyond these everyday annoyances, filament entanglement plays a fundamental role across a vast range of scientific and engineering contexts. At the macroscale, tangling governs the mechanical behavior of ropes, cables, and woven fabrics \cite{mclaren2006design,warren2018clothes}. At the molecular scale, the entanglement of flexible polymer chains governs the viscoelastic response of polymer melts and concentrated solutions and contributes significantly to the elastic behavior of polymer networks \cite{rubinstein2003polymer}. In biological systems, the entanglement of DNA within the cell nucleus, the knotting of proteins during folding, and the organization of collagen fibrils all depend critically on filament geometry\cite{arsuaga2002knotting,huang2005computer,perumal2008collagen}.

The mathematical study of knots has a long history, predating the influential practical compendium of Ashley\cite{ashley1944ashley} by well over a century: the systematic mathematical treatment of knot topology dates to Gauss and was substantially developed in the 19th and 20th centuries \cite{przytycki1998classical}. The statistical mechanics of spontaneous knot formation was subsequently developed using self-avoiding random walk models, which showed that longer walks increase the probability of knot formation\cite{frisch1961chemical,sumners1988knots} and that knot complexity scales with chain length\cite{shimamura2002knot}. More recently, macroscopic experiments on agitated strings in a rotating box\cite{raymer2007spontaneous,gendron2021time} and tumbled circular chains\cite{soh2019self} have revealed clear statistical relationships between filament length, confinement, and tangle complexity, connecting the mathematical theory to physical observables.

Despite this body of work, the majority of experimental studies have focused on straight filaments. Yet intrinsic curvature is a fundamental geometric feature of many natural and engineered filamentary systems: plant tendrils and climbing vines,\cite{mahadevan2000motility,goriely2006mechanics} curly hair,\cite{PhysRevLett.112.068103} and collagen fibrils,\cite{perumal2008collagen}, as well as active biological systems such as worms\cite{deblais2023worm}, where self-controlled tangling and untangling can emerge. How intrinsic curvature affects the spontaneous entanglement of macroscopic filaments under agitation, and how this effect varies with filament length, remain, to our knowledge, open questions.

Here, we present an experimental investigation of this question, focusing specifically on the role of natural curvature $\kappa$ and arc-length $L$ on the entanglement of soft elastic rods subjected to controlled bi-directional tumbling. This is an experimental exploration of a specific system, and the results reported here are specific to the geometric and dynamical parameters examined. We distinguish between two physically distinct entanglement modes that we define operationally: \textit{knotting} and \textit{locking}. We use the term \textit{knot} in a practical, rather than strictly mathematical, sense: a configuration is considered knotted if a self-entangled state persists upon pulling the two free ends of the open rod apart. In the strict mathematical definition, a knot requires the ends of the filament to be joined; our operational definition captures the physically relevant case of open rods and is consistent with prior macroscopic experimental studies\cite{raymer2007spontaneous}. \textit{Locking}, by contrast, refers to a configuration that prevents extraction of one end when pulled against a rigid constraint, regardless of whether it constitutes a topological knot. These two definitions allow consistent experimental quantification, and, as we demonstrate, the two modes exhibit different sensitivities to rod geometry under their respective testing protocols. 

\begin{figure}[h!]
    \centering
    \includegraphics[width=0.75\columnwidth]{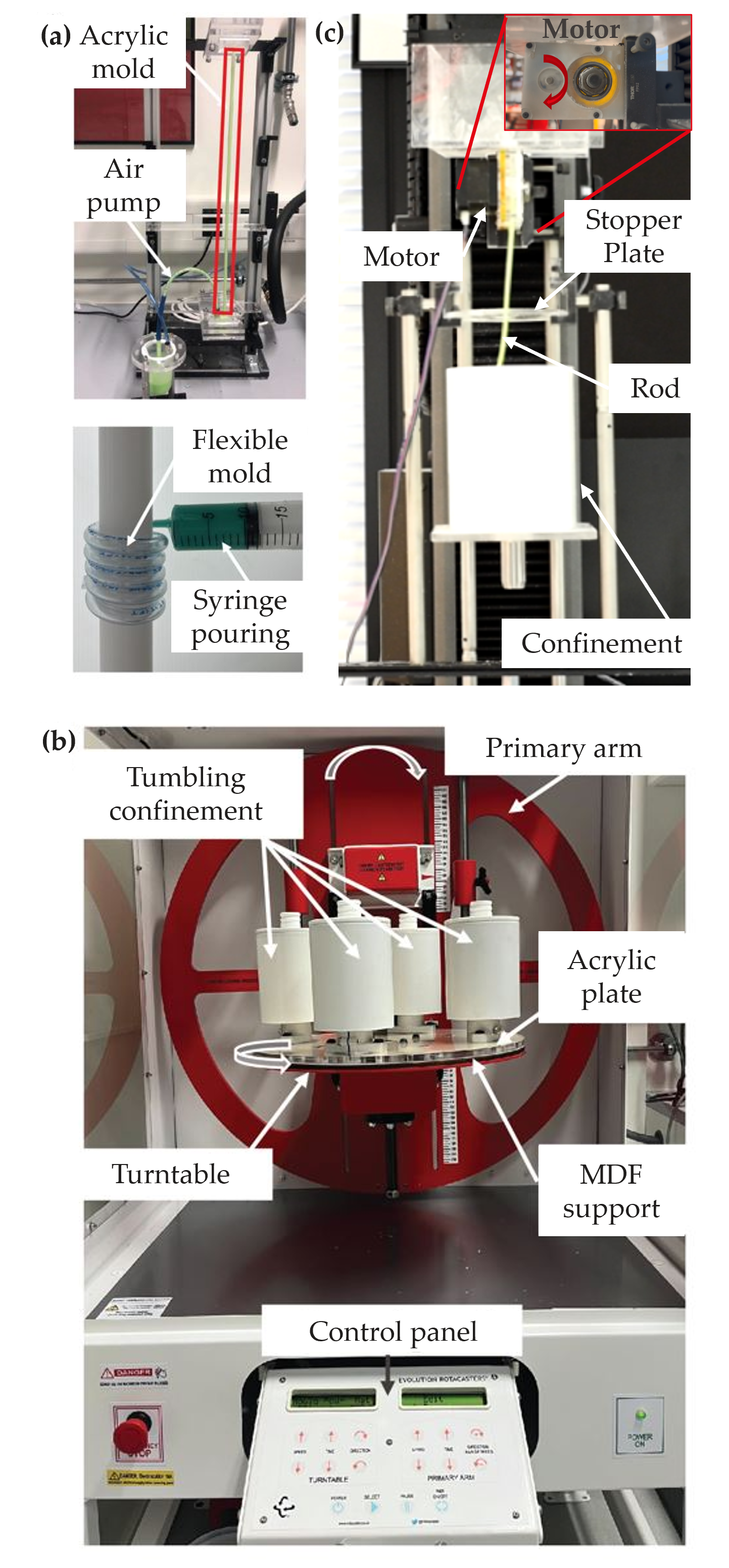}
    \caption{\textbf{Experimental overview.} (a)~Fabrication of straight rods via acrylic tube molds (top), and naturally curved rods via flexible PVC tube molds (bottom). (b)~Two-directional tumbling setup using the Evolution Rotacaster (rotational direction is denoted with arrows). (c) Mechanical testing setup via a stepper motor and the Instron machine.}
    \label{fig:setup}
\end{figure}

Elastic rods of diameter $d = 3.2~\mathrm{mm}$ were fabricated from vinylpolysiloxane (VPS-32, $E = 1.25~\mathrm{MPa}$, $\rho = 1170.2~\mathrm{kg/m^3}$, $\nu = 0.5$). Straight rods ($\kappa = 0$) were cast in rigid acrylic tubes \cite{grandgeorge2022elastic, johanns2021shapes, aymon2025self} (Fig.~\ref{fig:setup}(a)-top); naturally curved rods were cast in flexible PVC tubes wrapped around cylinders \cite{LAZARUS20131712, jawed2014coiling, khalid2014coiling} (Fig.~\ref{fig:setup}(a)-bottom) of outer radius $R_c \in [5,~45]~\mathrm{mm}$, yielding natural curvatures $\kappa = 1/(R_c + d/2) \in [0.02,~0.145]~\mathrm{mm^{-1}}$. Rod arc-lengths ranged from $L = 0.6$ to $4.5~\mathrm{m}$; longer rods were assembled by gluing shorter sections. All rods were coated with talc powder to reduce friction\cite{grandgeorge2021mechanics}. Rods were subjected to bi-directional tumbling (Supplementary Video 1) for 20~minutes\cite{gendron2021time} inside cylindrical confinements of inner diameter $D = 99~\mathrm{mm}$ and height $H = 130~\mathrm{mm}$ (Fig.~\ref{fig:setup}b), using an Evolution Rotacaster, with its turntable rotating about the vertical axis at $12.5~\mathrm{rpm}$ and its primary arm rotating about a horizontal axis at $7.5~\mathrm{rpm}$. All experiments were performed using this same pair of tumbling speeds; the possible influence of rate-dependence on the observed statistics was not examined here. Each experimental condition was repeated 15 times, yielding probabilities reported with $95\%$ Wald confidence intervals throughout.

\begin{figure}[h!]
    \centering
    \includegraphics[width=0.76\columnwidth]{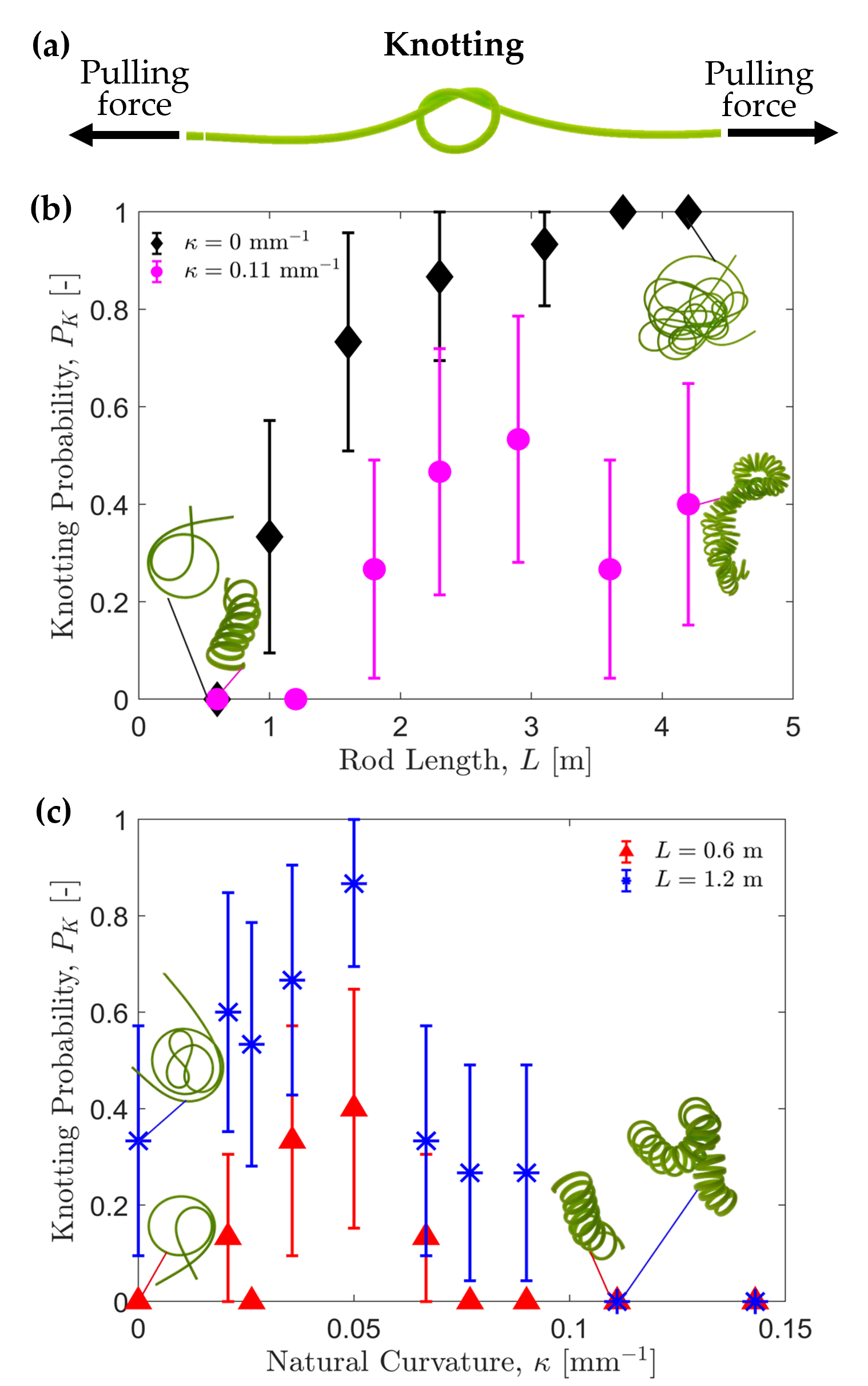}
    \caption{\textbf{Knotting results.} (a)~Knotting test: both free ends are pulled apart to detect self-entanglement. (b) Knotting probability~$P_K$ versus arc-length $L$ for straight ($\kappa = 0$) and highly curved ($\kappa = 0.11~\mathrm{mm^{-1}}$) rods. (c) Knotting probability~$P_K$ versus natural curvature $\kappa$ for rod lengths $L = 0.6~\mathrm{m}$ and $L = 1.2~\mathrm{m}$. Error bars indicate $95\%$ confidence intervals. }
    \label{fig:knotting}
\end{figure}

\begin{figure}[h!]
    \centering
    \includegraphics[width=0.76\columnwidth]{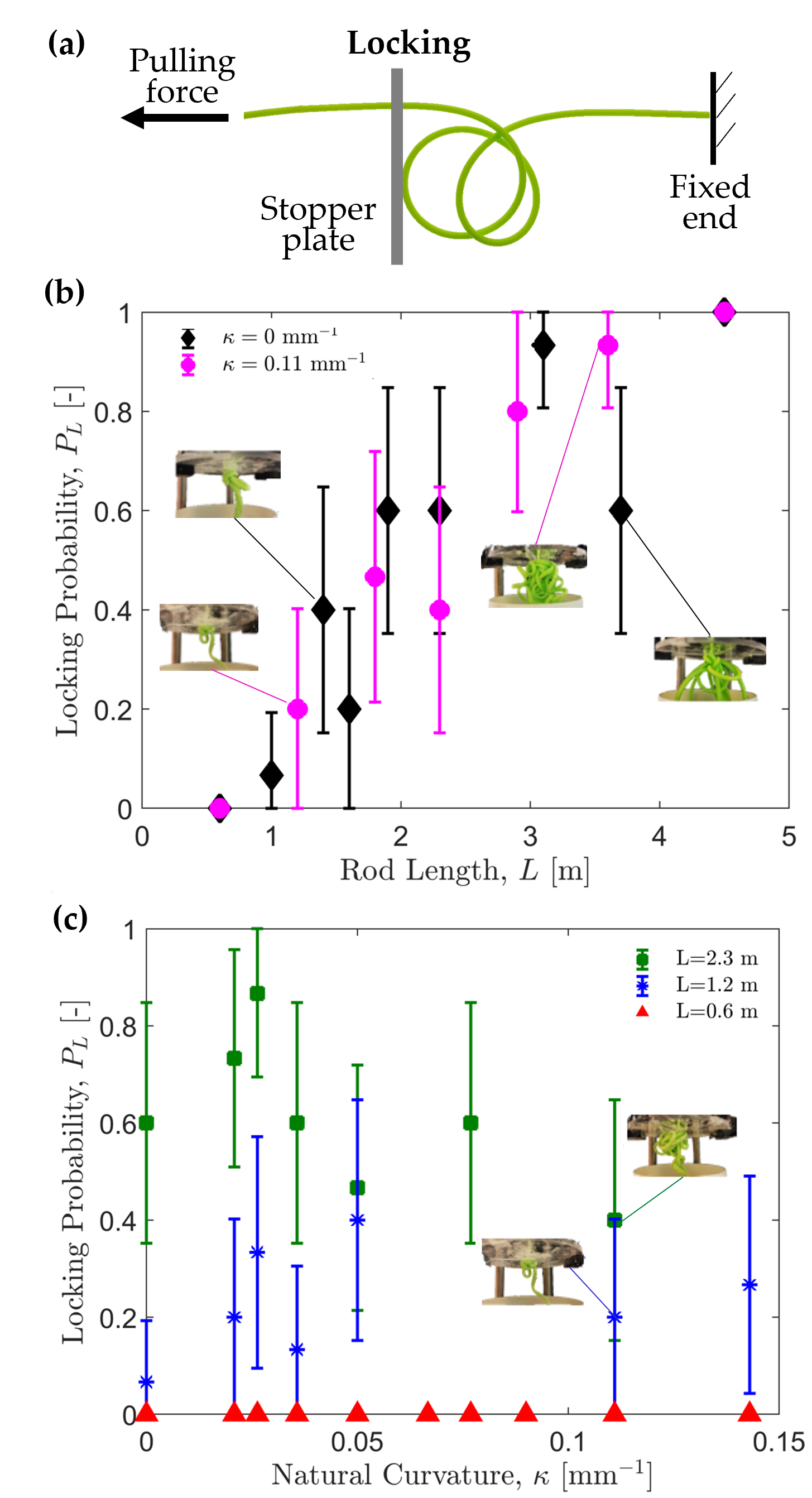}
    \caption{\textbf{Locking results.} (a)~Locking test: one end is fixed, the other pulled against an acrylic stopper plate via a stepper motor mounted in an Instron machine. (b) Locking probability~$P_L$ versus arc-length $L$ for straight ($\kappa = 0$) and highly curved ($\kappa = 0.11~\mathrm{mm^{-1}}$) rods. (c) Locking probability $P_L$ versus natural curvature $\kappa$ for rod lengths $L = 0.6~\mathrm{m}$, $L = 1.2~\mathrm{m}$ and $L = 2.3~\mathrm{m}$. Error bars indicate $95\%$ confidence intervals.}
    \label{fig:locking}
\end{figure}

Knotting probability, $P_K$, is defined as the fraction of trials in which, after tumbling with both ends free, a persistent self-entanglement is detected upon pulling both ends apart (Fig.~\ref{fig:knotting}a). The locking probability, $P_L$, is defined as the fraction of trials in which the rod, after tumbling with both ends fixed, resists extraction when one end is pulled against the stopper plate (Fig.~\ref{fig:locking}a). Pulling was imposed at a constant linear translation speed $v_\mathrm{linear} = 0.1615~\mathrm{m.s^{-1}}$ by a stepper motor (Quimat 17HS13-0104S), while the corresponding pulling force was recorded by the load cell of an Instron universal testing machine (Fig.~\ref{fig:setup}c; Supplementary Video 2). 

Fig.~\ref{fig:knotting}(b) shows $P_K$ as a function of rod arc-length $L$ for two extreme values of natural curvature. For straight rods ($\kappa = 0$), $P_K$ increases monotonically with $L$, reaching $P_K = 1$ for rods longer than $3.5~\mathrm{m}$, consistent with prior observations on tumbled strings.\cite{raymer2007spontaneous} In contrast, rods with $\kappa = 0.11~\mathrm{mm^{-1}}$ exhibit systematically lower $P_K$ across all lengths, by as much as 73\%. Remarkably, $P_K$ decreases at $L \approx 3.7~\mathrm{m}$ rather than continuing to rise, suggesting that the helix becomes geometrically jammed or arrested within the cylindrical confinement at this length. Fig.~\ref{fig:knotting}(c) reveals that the dependence of $P_K$ on $\kappa$ appears non-monotonic for fixed rod lengths. For both $L = 0.6~\mathrm{m}$ and $L = 1.2~\mathrm{m}$, $P_K$ first increases near $\kappa \approx 0.05~\mathrm{mm^{-1}}$ before decreasing to zero at a critical curvature $\kappa^* \approx 0.11~\mathrm{mm^{-1}}$, beyond which knotting is fully suppressed within our experimental ensemble. The physical origin of this non-monotonic behavior is not yet fully understood and remains an open question. These trends may be specific to the rod diameter, confinement geometry, and tumbling conditions studied here, and their generality to other systems should be investigated in future work.

Turning to locking, Fig.~\ref{fig:locking}(b) shows that the locking probability $P_L$ increases with rod length in a manner that is essentially independent of natural curvature across the full range of $\kappa$ explored. In Fig.~\ref{fig:locking}(c), for $L = 2.3~\mathrm{m}$, we find $P_L \approx 0.6$ almost regardless of $\kappa$, while rods with $L = 0.6~\mathrm{m}$ never produced a lock event across all 15 repetitions. This insensitivity of $P_L$ to curvature contrasts sharply with the strong geometric control observed for $P_K$, and suggests that knotting and locking are governed by physically distinct mechanisms: knotting is sensitive to global rod geometry, while locking appears to be driven primarily by local confinement-induced contact and loop formation, which are less affected by the overall shape of the rod.

\begin{figure}[b!]
    \centering
    \includegraphics[width=0.85\columnwidth]{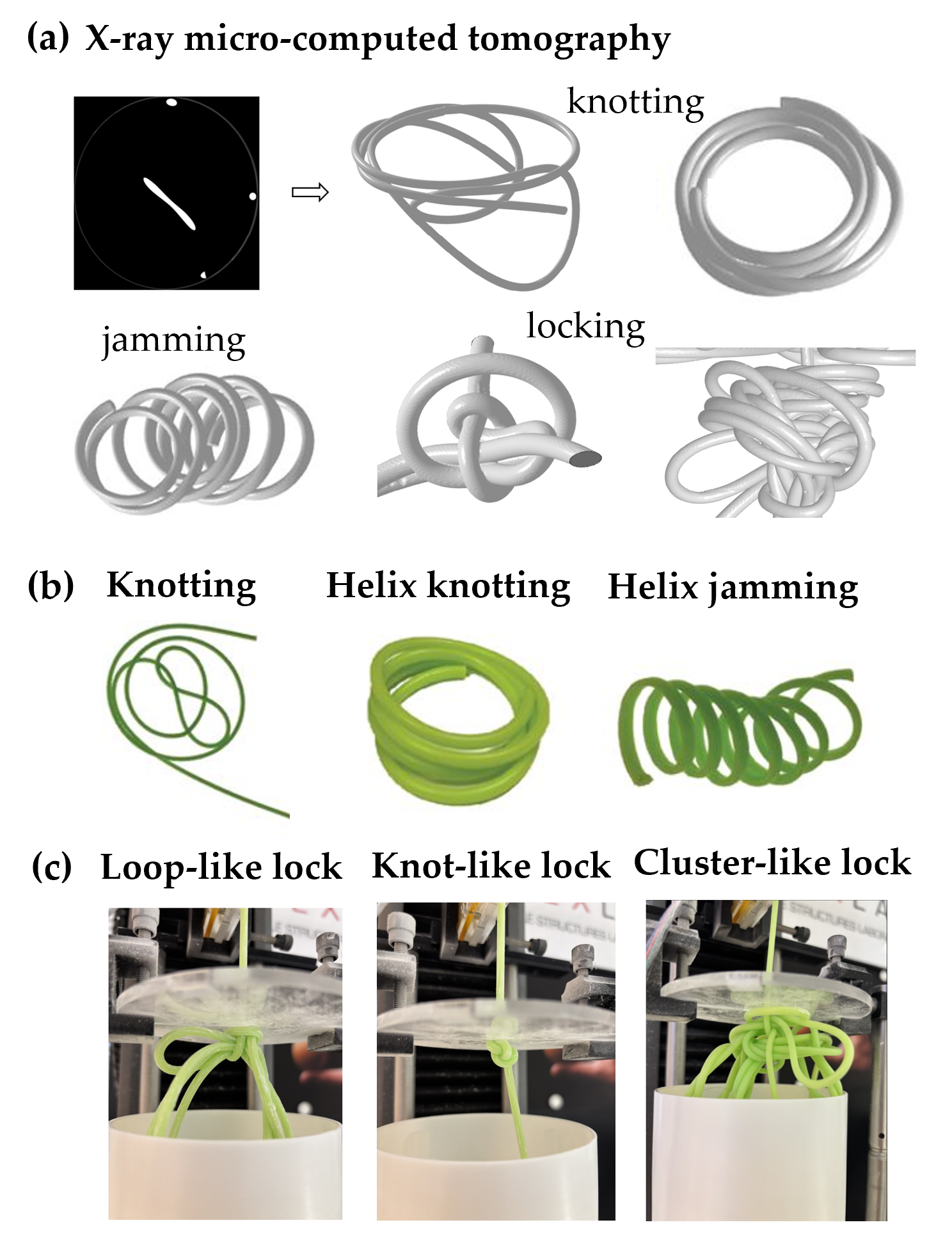}
    \caption{\textbf{Knotting and locking complexity.} 
(a)~X-ray micro-computed ($\mu$CT) tomography and centerline extraction for visualization of selected entangled configurations. 
(b)~Representative configurations for a straight rod (left), curved rod with $\kappa = 0.05~\mathrm{mm^{-1}}$ (middle), and highly curved rod with $\kappa = 0.11~\mathrm{mm^{-1}}$ (right). 
(c)~Three recurring locking morphologies: loop-like (left), knot-like (centre), and cluster-like (right).}
    \label{fig:complexity}
\end{figure}

To characterize the three-dimensional structure of selected entangled configurations, we performed X-ray micro-computed tomography ($\mu$CT, Scanco Medical $\mu$CT~100, voxel size $34.2~\mu\mathrm{m}$%
) on selected samples (Fig.~\ref{fig:complexity}a) and extracted rod centerlines in MATLAB.\cite{grandgeorge2021mechanics,grandgeorge2022elastic,johanns2021shapes} The tomographic reconstructions revealed internal rod configurations that were difficult to resolve from external photographs alone, while the extracted centerlines aided visualization of the three-dimensional geometry. Straight rods form highly entangled structures with large and variable, geometrically complex configurations, while naturally curved rods consistently yield simpler entangled structures across the explored parameter space, consistent with the suppression of knotting at higher curvature. Fig.~\ref{fig:complexity}(b) illustrates this contrast with representative configurations. This reduction may be associated with helix jamming during tumbling. These simpler structures most often resemble overhand or figure-eight knots, in contrast to the disordered configurations observed in longer straight rods, suggesting a geometric simplification of entanglement with increasing curvature.

Three recurring lock morphologies were identified (Fig.~\ref{fig:complexity}c), characterized via $\mu$CT imaging of selected configurations. \textit{Loop-like} locks consist of configurations in which the rod forms one or more loops around itself before tightening against the stopper plate, and were most commonly observed at medium rod lengths. This morphology resembles the locking unit identified in the bowline knot \cite{aymon2025self}. \textit{Knot-like} locks are tight, compact configurations that resist untangling; they resemble overhand or figure-eight knots and were the dominant morphology in shorter rods for both straight and helical geometries. \textit{Cluster-like} locks are highly disordered, multi-contact configurations that formed rapidly during pull-out tests and dominated in longer rods; these structures were difficult to characterize topologically and to reproducibly untangle. Across all three morphologies, the onset of locking occurred at different stages of the pull-out process, with no consistent relationship observed between locking onset and the axial position of the tangled region along the rod. 

Taken together, our results reveal a clear mechanistic distinction between knotting and locking in tumbled elastic rods. Knotting is strongly governed by the global geometry of the rod: natural curvature reduces the effective span of the configuration and limits end-to-end crossing events, ultimately suppressing knot formation above a critical curvature. Locking, by contrast, appears to be governed primarily by local confinement-induced contact and loop formation, and is largely insensitive to intrinsic curvature. This separation of mechanisms suggests that the two entanglement modes can, in principle, be independently controlled through geometric design. Within the specific experimental parameter space studied here, no knotting events were observed for rods with sufficiently high natural curvature, while their locking behavior remained controlled primarily by rod length. We note that generalization of these findings to other rod diameters, materials, confinement geometries, or tumbling speeds would require further systematic investigation.

Although this work focuses on macroscopic elastic rods, we speculate that similar geometric principles may play a role in the entanglement of naturally curved biological filaments at the microscale---such as helical proteins and supercoiled DNA\cite{arsuaga2002knotting,tubiana2013spontaneous}, where intrinsic curvature similarly shapes the effective geometry of the chain. A direct quantitative connection would require dedicated theoretical and computational modeling, which we leave for future work.

\section*{Conflicts of interest}
There are no conflicts to declare.\\

\section*{Data availability}
The supplementary experimental videos are provided in the Supplementary Information.

\section*{Acknowledgements}
We thank Jorn Dunkel and Vishal Patil for fruitful discussions, and Aymeric Rio and Elie Saliba for proof-of-concept experiments.

\bibliographystyle{apssamp}
\bibliography{apssamp}

\end{document}